\documentclass[a4paper]{jpconf}
\usepackage{graphicx}
\begin{document}
\title{Moving vortex matter with coexisting vortices and anti-vortices}

\author{Gilson Carneiro}

\address{Instituto de F\'{\i}sica, Universidade Federal do Rio de Janeiro 
C.P. 68528,  21941-972 Rio de Janeiro-RJ, Brazil}

\ead{gmc@if.ufrj.br}

\begin{abstract}
Moving vortex matter, driven by  transport  currents independent of time, in which vortices and anti-vortices coexist is  investigated  theoretically in thin superconducting films with nanostructured defects. A simple London model is proposed for the vortex dynamics in  films with periodic  arrays of nanomagnets or  cylindrical holes (antidots). Common to these films is that vortex anti-vortex pairs may be created in the vicinity of the defects by relatively small transport currents,  because it adds to the  current generated by the defects - the nanomagnets screening current, or the antidots backflow current - and may exceed locally the critical value for vortex anti-vortex pair creation. The model assumes that    vortex matter dynamics is governed by Langevin equations, modified to account for creation and annihilation of vortex anti-vortex pairs. For pair creation, it is assumed that whenever the total current at some location exceeds a critical value, equal to that needed to separate a vortex from an anti-vortex by a vortex core diameter, a pair is created instantaneously around this location. Pair annihilation occurs by vortex  anti-vortex collisions. The model is applied  to films at zero external magnetic field and low temperatures. It is found that several moving vortex matter steady-states with equal numbers of vortices and anti-vortices are possible. 
\end{abstract}
The dynamics of vortex matter in thin superconducting films with nanostructured defects has attracted a great deal of attention recently \cite{rev}. Most works done so far assume that the vortex matter is made up of identical vortices with unit vorticity,  and  that the laws governing its motion are those for ordinary matter consisting of identical classical particles with zero mass \cite{dyn}. However, an important difference between vortex matter and  ordinary matter is that vortex anti-vortex (v-av) pairs can be created by thermal fluctuations or by supercurrents, and annihilated by collisions \cite{vav}. These processes can be neglected only at  low temperatures, and for  supercurrents everywhere small compared to the critical current for v-av pair creation. These conditions are well satisfied in homogeneous  films at low temperatures, and when the transport current is uniform, since it cannot exceed the critical current, otherwise superconductivity is destroyed by proliferation of v-av pairs.  However, in  films with nanostructured defects, the supercurrent is not in general uniform, and may exceed the critical current locally, creating only a few  v-av pairs. As a consequence, superconductivity is not destroyed,  and  states of moving vortex matter with coexisting vortices and anti-vortices  are produced. These states  exist even in the absence of applied magnetic fields. Two systems are considered here: i) films with arrays of nanomagnets, for which the transport current is uniform, but a non-uniform screening current flows, induced by the nanomagnets. ii) films with arrays of holes, for which the transport current itself is non-uniform, because it cannot go through the holes.  This paper demonstrates the existence of such states at zero applied fields, and studies  some of its properties. 

\begin{figure}[b]
\centerline{\includegraphics[scale=0.2]{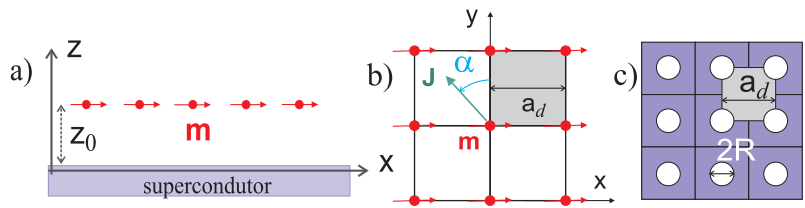}}
\caption{ Film and defect  arrays: a) Side view of dipole array. b) Top view of dipole array, and transport current orientation. c) Top view of antidot array. }
\label{fig.f1}
\end{figure}

The theoretical model is the following. Vortex matter in thin superconducting films, characterized by   thickness $d$, penetration depth $\lambda\gg d$, and vortex core radius  $\xi$, interact with the following defect arrays: i) a square lattice of  magnetic dipoles, with magnetic moment ${\bf m}$ parallel to one of the unit cell sides (${\bf m}=m\hat{\bf x}$ ), placed  at a distance $z_0$ from the film  surface ( Fig.\ref{fig.f1}). ii) a square lattice of cylindrical holes (antidots) of radius $R\gg \xi$ ( Fig.\ref{fig.f1}). In both cases the defect lattice unit cell has dimensions $a_d\times a_d$. 
%
%
The  vortex matter is modeled as a system of classical particles with zero mass. To account for v-av pair creation and annihilation, it is assumed that {\it the time evolution can be broken into two  parts: motion of the vortices and anti-vortices, and  creation of v-av pairs}. The first part is governed by Langevin equations \cite{dyn}
\begin{equation}
\eta \frac{d{\bf r}_j(t)}{dt} = q_j( {\bf F}^{v}_j + {\bf F}^{d}_j+ {\bf F}_j) +{\mbox{\boldmath $\Gamma$}_j} \;,
\label{eq.lgv} 
\end{equation} 
where $\eta$ is the friction coefficient, $j$ labels  vortices and anti-vortices present at time $t$, 
${\bf r}_j(t)$  and $q_j=\pm 1, \pm 2, ...$ denote,  respectively, the corresponding  position  vectors and vorticities  ($q>0$ for vortices, $q<0$ for anti-vortices).   ${\bf F}^{v}_j$, and  ${\bf F}^{d}_j$ are, respectively,  the forces of interaction of  a test vortex ($q=1$) located at  ${\bf r}_j(t)$ with   the other vortices and anti-vortices, and  with the defect array. ${\bf F}_j$  is  the Lorentz force of the transport current, ${\bf J}$,  on a test vortex at  ${\bf r}_j(t)$, that is ${\bf F}_j=(\phi_0 d/c){\bf J}\times \hat{\bf z}$. ${\mbox{\boldmath $\Gamma$}_j}$ is the random force.  In the  film with the dipole array,  ${\bf J}$ is uniform  and makes an angle $\alpha$ with the $y$-axis ( Fig.\ref{fig.f1}). In both cases ${\bf J}$ is independent of time. The forces ${\bf F}^{v}$ and ${\bf F}^{d}$ are also the Lorentz forces on a test vortex due, respectively,  to the  currents generated by the  other vortices and anti-vortices, ${\bf J}_{v}({\bf r})$, and by the defect array, ${\bf J}_{d}({\bf r})$.  These equations naturally account for v-av pair annihilation,  which happen when a vortex and an anti-vortex collide. For  v-av pair creation, {\it the model assumes that if,  at a given  instant of time, the current at some location exceeds a critical value, a v-av pair is added around this location instantaneously}. 

The  force ${\bf F}^{v}$ is given by ${\bf F}^{v}_j = -{\mbox{\boldmath $\nabla$}}_j\sum_i q_i U_{vv}({\bf r}_j-{\bf r}_i)$, where $U_{vv}({\bf r})$ is the  vortex-vortex interaction potential in the London limit. Essentially, $U_{vv}({\bf r})\sim -\epsilon_0d\ln{(r/\Lambda)}$ ($\epsilon_0=(\phi_0/4\pi\lambda)^2$, and  $\Lambda=2\lambda^2/d$), for $r>\xi$. For $r<\xi$, a Gaussian cutoff  is used in $U_{vv}$, in order to account for the vortex core \cite{ehb}.  Creation of a v-av pair by a supercurrent occurs if its Lorentz force can separate a vortex from an anti-vortex by a distance $\sim 2\xi$, along the direction of the Lorentz force.  The critical current, $J_m$, is  estimated by balancing the energy change for pair creation, $\Delta E= 2E_{sf}+ 2U_{vv}(2\xi)$ ($E_{sf}\sim \epsilon_0d\ln{(\xi/\Lambda)}=$ vortex self energy), with the work done by the Lorentz force, $2\xi \phi_0d\,J_m/c$. The result is $J_{m} \sim (c\epsilon_0/\phi_0 \xi)\;\ln{2}= 0.9J_d$, where $J_d=c\phi_0/(12\sqrt{3}\,\pi^2\lambda^2\xi)$ is the depairing current.  The relevant current for v-av pair creation is the total current, ${\bf J}_T={\bf J}+{\bf J}_{d}+{\bf J}_{v}$. When  ${\bf J}_T$  exceeds $J_m$ locally,  v-av pairs are created. The detailed procedure for v-av pair creation  is  discussed next for each  defect array. 
%

\begin{figure}[b]
\centerline{\includegraphics[scale=0.2]{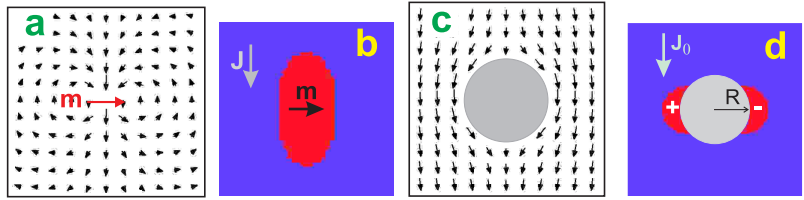}}
\vspace{5mm}
\caption{ a) Screening current around a dipole (arrow). b) Regions  below a dipole where  v-av pairs are created (red), for $\alpha=135^o$, $J/J_m=0.55$. c) Transport current in the vicinity of a circular hole of radius $R$ for 
${\bf J}_0$ in the negative $y$-direction. d) Regions near the hole where a vortex (red $+$) or anti-vortex ( red $-$ ) are created, for $J/J_m=0.75$. } 
\label{fig.f2} 
\end{figure}

\noindent{\bf  Dipole Array}. 
The vortex-defect force is  the magnetic interaction between the vortex and the dipoles,  given by  ${\bf F}^{d}({\bf r})=-{\mbox{\boldmath $\nabla$}\sum_{{\bf R}_d}}{\cal{U}}({\bf r}-{\bf R}_d)$ where  ${\cal{U}}({\bf r})=(\epsilon_0d)\;(4\pi m/\phi_0 z_0)\;(z_0{\bf r}\cdot\hat{\bf x}/r^2)(1-(1+r^2/z^2_0)^{-1/2})$ is the potential of a single dipole,  ${\bf r}$ is the vortex position vector with respect to the dipole position (projected in the film plane), and  ${\bf R}_d$ denotes the dipole array translation vectors.  The  screening current ${\bf J}_{d}$ follows  from ${\bf F}^{d}=(\phi_0 d/c){\bf J}_d\times \hat{\bf z}$. The resulting ${\bf J}_d $ for a dipole array with  $a_d\gg z_0$ is shown in  Fig.\ref{fig.f2}a, close to the position of one of the dipoles, where $J_d$ is the largest. The following  procedure is used to  add v-av pairs. If at a point ${\bf r}_0$, ${\bf J}_T({\bf r}_0)$ exceeds $J_m$ ($ J_T({\bf r}_0) > J_m$),  a v-av pair is added around ${\bf r}_0$:  a vortex ($q=1$) at ${\bf r}_0 +{\mbox{\boldmath $\delta$}}$, and an anti-vortex  at ${\bf r}_0 -{\mbox{\boldmath $\delta$}}$, where ${\mbox{\boldmath $\delta$}}$ has modulus  $\xi$ and is oriented parallel to  the Lorentz force of  ${\bf J}_T({\bf r}_0)$, that is ${\mbox{\boldmath $\delta$}}=\xi {\bf J}_{T}({\bf r}_0)\times \hat{\bf z}/ J_{T}({\bf r}_0)$. A fundamental assumption of the model is that {\it v-av pairs can be added in the manner described above only around points ${\bf r}_0$ for which} $\mid {\bf J}+{\bf J}_{d}({\bf r}_0)\mid > J_m$. These points are localized  in regions close  to the dipole positions, or between them, depending the orientation and magnitude of ${\bf J}$. An example is shown in Fig.\ref{fig.f2}b. This assumption is necessary in order to avoid unphysical behavior caused by the vortex core. What happens is that the supercurrent generated by a  vortex or an anti-vortex  has values  close to $J_m$ in the proximity of the core. When this current adds to another -  transport, screening, or from another vortex or anti-vortex - the resultant may exceed $J_m$ in the proximity of the core, leading to creation of v-av pairs. This is unphysical. What should happen in a more realistic model is  deformation of the vortex core.  A more detailed  justification of this assumption will be given elsewhere.
%
%

\noindent{\bf   Antidot Array}. Here only a single antidot is considered in detail, which is relevant for arrays  where the antidots  are far apart ($a_d>>R \gg \xi$). A vortex, with vorticity $q$, interacts with the antidot through the two images it induces on the hole: one, with vorticity $q$, at the center of the hole (${\bf r}=0$), the other, with vorticity $-q$,  at ${\bf r}_I= (R/r)^2\;{\bf r}$, where  ${\bf r}$ is the vortex position vector \cite{cho}. Here this interaction is included in  ${\bf F}^{v}$, since it depends on the vortex matter distribution. Thus, ${\bf J}_{v}$ also contains  the current generated by the vortex images outside the holes, and ${\bf F}^{d}=0$. The  transport current is the sum of  an uniform  current ${\bf J}_0$, with the  backflow current ${\bf J}_{bf}$, that is  ${\bf J}={\bf J}_0+{\bf J}_{bf}$.  For the  circular hole, ${\bf J}_{bf}$ has a dipolar spatial dependence, identical to that of an ideal fluid.  An example of ${\bf J}$ is shown  in 
Fig.\ref{fig.f2}c. The maximum values of $J$ occurs at two points located in the circumference of the  hole  perpendicular to ${\bf J}_0$. At these points  $J=2J_0$. Thus, when $J_0>=0.5J_m$, $J$ exceeds $J_m$ in regions located in the vicinity of these points,  and v-av pairs are  created \cite{he4}. A vortex (anti-vortex) is added  in the part of the region where $J>J_m$ and the Lorentz force on the vortex (anti-vortex) points away from the hole center. An example of these regions is shown in Fig.\ref{fig.f2}d. The critical current for  for v-av pair creation is, essentially, $J_m$ in the limit $R\gg \xi$. Thus, the final result of this process is creation of a v-av pair near the defect, as in the film with the dipole array. 
%
\begin{figure}[t]
\centerline{\includegraphics[scale=0.2]{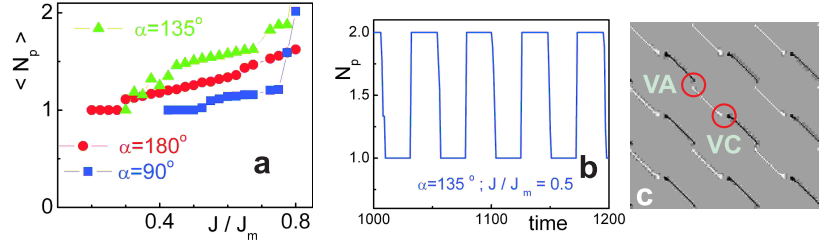}}
\vspace{5mm}
\caption{ Results for a film with dipole array ($m=0.85\phi_0\xi,\; z_0=2\xi,\; a_m=16\xi$). a) Time averaged number of vortices per dipole vs. transport current. b) Number of vortices per dipole vs. time (arbitrary units). c) Trajectories of vortices (light) and anti-vortices (dark). Circles: regions where v-av pairs are created (VC) and annihilated (VA). }
\label{fig.f3}
\end{figure}
%
%

Numerical solutions  of Eq.\ (\ref{eq.lgv}) are obtained  at low temperatures for the  film with the dipole array, using a discrete version of  Eq.\ (\ref{eq.lgv}) on a square grid. The results confirm the existence of a rich variety of steady-states with equal numbers of vortices and anti-vortices. Typical ones are shown in Fig.\ref{fig.f3}, for  $J<J_m$, $90^o<\alpha<180^o$, and  
$m$  chosen so that the screening current is everywhere smaller that $J_m$.  The number of v-av pairs per dipole, $N_p$, averaged over time (denoted $<N_p>$) depends on $J$ and $\alpha$ as shown in Fig.\ref{fig.f3}a.   The time dependence of $N_p$, shown in Fig.\ref{fig.f3}b,  is oscillatory, varying between one and two,  almost  periodically  in this particular case. Typical trajectories for the vortices and anti-vortices and the regions where v-av pairs are created  and annihilated are  shown in Fig.\ref{fig.f3}c. For $\alpha<80^o$ no steady-states are found for $J<J_m$ \cite{ccg}. The  behavior shown in Fig.\ref{fig.f3} is  a direct consequence of the model basic assumptions. The  dependence of $<N_p>$ on $\alpha$ results from the spatial dependence of  the screening current (Fig.\ref{fig.f2}a): for  $\alpha>90^o$ the transport current and the screening current add near dipoles, favoring v-av pair creation, whereas for $\alpha<90^o$ the opposite occurs. The time-dependence of $N_p$ is due to oscillations in time of $J_T$, caused by the motion of the vortex matter. During the time interval when $N_p(t)=1$, $J_T< J_m$ in the regions where v-av pairs are created, but increases in these regions  as the  vortices and the anti-vortices move away from the dipoles. Eventually $J_T>J_m$ in parts of  these regions and one new  v-av pairs is created by each dipole. After that  $N_p(t)=2$, until collisions between  vortices  and  anti-vortices, taking place  between the dipoles (Fig.\ref{fig.f3}c),  reduce it to one. 

Steady-states of moving vortex matter with similar properties are expected to exist in films with periodic arrays of holes, as well as in films with non-periodic defect arrays.

\ack
Research supported in part by the Brazilian National Research Council, CNPq.
\section{References}
%

\end{document}